\documentclass[12pt]{article}
\usepackage{amssymb,amsthm,amsxtra}
\newtheorem{thm}{Theorem} 
\theoremstyle{remark}
\newtheorem*{example}{Example}
\theoremstyle{definition}
\newtheorem{defi}{Definition} 
\newcommand{\C}{\ensuremath{\mathbb{C}}} 
\newcommand{\CN}{\ensuremath{\mathbb{C}^{N}}} 
\newcommand{\R}{\ensuremath{\mathbb{R}}}  
\newcommand{\Eins}{\ensuremath{\mathbf{1}}}

\newcommand{\Cinfty}{\ensuremath{\mathcal{C}^{\infty}}} 
\newcommand{\Dpr}[1]{\ensuremath{\mathcal{D}'(#1)}}
\newcommand{\supp}{\mathrm{supp}}
\newcommand{\singsupp}{\mathrm{sing\,supp\,}} 
\newcommand{\ch}{\ensuremath{\mathrm{char}}}
\newcommand{\WF}[1]{\ensuremath{W\!F(#1)}}
\newcommand{\WFpol}[1]{\ensuremath{W\!F_{pol}(#1)}}
\newcommand{\Mf}{\ensuremath{\mathcal{M}\/}}
\newcommand{\TM}{\ensuremath{T\Mf}\/} 
\newcommand{\TastM}{\ensuremath{T^\ast\!\Mf}} 
\newcommand{\TastMon}{\ensuremath{\TastM\setminus \mathbf{0}}} 
\newcommand{\TastMMon}{\ensuremath{T^\ast\!(\Mf\times\Mf)\setminus \mathbf{0}}} 
\newcommand{\DM}{\ensuremath{D\Mf}\/} 
\newcommand{\DastM}{\ensuremath{D^\ast\!\Mf}} 
\renewcommand{\slash}[1]{\mbox{\ensuremath{\not\!#1}\/}}
\newcommand{\xislash}{\slash{\xi}}
\newcommand{\nslash}{\slash{\nabla}}
\begin{document}
\title{
{\normalsize March 2000\hfill DESY 00--043\\ \hfill math-ph/0003015\\[2.5cm]}
{\bf Singularity structure of the\\ two point function of the free Dirac field\\
on a globally hyperbolic spacetime}\\[1cm]}
\author{Kai Kratzert\\[2mm]
{\normalsize\it Deutsches Elektronen-Synchrotron DESY}\\ 
{\normalsize\it Notkestrasse 85, D--22603 Hamburg, Germany}\\[2mm]
{\normalsize\tt kai.kratzert@desy.de}\\[6mm]}
\date{March 10, 2000\\[2cm]}
\maketitle


\begin{abstract}
  We give an introduction to the techniques from microlocal analysis
  that have successfully been applied in the investigation of Hadamard
  states of free quantum field theories on curved spacetimes. The
  calculation of the wave front set of the two point function of the
  free Klein-Gordon field in a Hadamard state is reviewed, and the
  polarization set of a Hadamard two point function of the free Dirac
  field on a curved spacetime is calculated.
\end{abstract}
\newpage

\section{Introduction}

The general framework of this work is that of quantum field theory on
curved spacetime. Since we still do not have a full theory of quantum
gravity, as a first step we try to include gravitational effects into
the quantum field theoretical description of our world by a
semi-classical approximation. The gravitational field is considered as
a fixed classical background spacetime on which matter is described by
a quantum field theory.  Since the Planck length, at which effects
from quantum gravity are expected to become important, is very small,
this approximation should have a wide range of validity and should be
appropriate to predict quantum effects near black holes or in the
early universe.  One famous result in this framework is the Hawking
radiation of black holes.

The main problem in the quantization of a field theory on a curved
background is that one cannot make use of Poincar\'e invariance like
on Minkowski spacetime, where this symmetry fixes a unique vacuum
state. On a curved spacetime, however, there is no such natural choice
of a state, and the central task is to single out the physically
acceptable ones.

For free, linear models it has turned out that these acceptable states
are the quasi-free Hadamard states. They are determined by their two
point function which has got a prescribed singularity structure fixed
by geometry.  For the free Klein-Gordon field it was recently
discovered by Radzikowski that this Hadamard condition, which at first
sight is a global condition, can be reformulated as a local condition
in terms of wave front sets.  This allows one to apply the elegant and
powerful mathematical techniques from microlocal analysis. In this way
it could be proved that certain states known for a long time are
indeed Hadamard states. A microlocal spectrum condition for
interacting quantum field theories was formulated and used to define
Wick polynomials and to prove the renormalizability of interacting
scalar field theories on globally hyperbolic spacetimes.

For more realistic quantum field theories, however, first of all the
Dirac field, more or less nothing was done, not because of conceptual
problems, but because of the greater complexity due to the
multi-component nature of the Dirac field.  In this work the
singularity structure of the two point function of the free Dirac
field will be investigated. We will make use of the polarization set
which was introduced by Dencker in order to analyze the singularities
of vector-valued distributions. Applying Dencker's results, we will
show that the singularities of solutions of the Dirac equation
propagate in a simple way.  This will be used to calculate the
polarization set of the two point function for Hadamard states of the
Dirac field. Our analysis makes use of unpublished results by
Radzikowski \cite{radzikowski_unp}, and our results are in agreement with those
independently obtained by Hollands \cite{hollands} using a slightly
different approach.

This paper is organized as follows: The first chapter after the
introduction gives an overview of the physical problems we are
interested in and fixes the notation used throughout this work. We
introduce the Klein-Gordon and Dirac fields on a curved spacetime and
give a definition of Hadamard states. After that, we come to the
mathematics and review the methods from microlocal analysis on which
our analysis is based. The central definitions are the wave front and
polarization sets, and the main results are the theorems on the
propagation of singularities. As a side result, we will rederive the
well-known properties of light propagation in a curved spacetime in a
mathematically very elegant way.  In the fourth chapter these methods
are applied to our physical objects, and we state our main results.
Finally we give an outlook and an overview about possible applications
of our results.


\section{Quantum field theory on curved spacetime}

\subsection{Notions from general relativity}

In the general theory of relativity, for an introduction see
\cite{wald_gr}, spacetime is represented by a differentiable manifold
$(\Mf,g)$ equipped with a Lorentzian metric $g$, i.e.\ a
differentiable, symmetric, and non-degenerate type $(0,2)$ tensor
field of signature $(+ - - -)$. For simplicity, we restrict ourselves
to the physical four dimensions.

As usual, the tangent bundle will be denoted \TM, and the cotangent
bundle is \TastM.  For the set $\{(x,\xi)\in\TastM\,|\,\xi\neq 0\}$ we
will shortly write \TastMon. The canonical projection will always be
denoted $\pi:\TastM\rightarrow\Mf$. Any vector bundle $E$ over \Mf\/
can be pulled back to a vector bundle $\pi^\ast E$ over \TastM.  The
set of smooth sections of a vector bundle $E$ over \Mf\/ will be
denoted $\Gamma(\Mf,E)$, sections having compact support form the set
$\Gamma_0(\Mf,E)$.

The Levi-Civit\'a connection is the unique connection on $\TM$ that is
symmetric and compatible with the metric. The associated covariant
derivative is denoted $\nabla$. The Riemann tensor is the curvature
tensor of this connection which is contracted to the Ricci tensor and
finally to the scalar curvature $R$.

We will restrict ourselves to globally hyperbolic spacetimes. These
are spacetimes that admit a Cauchy surface, that is a spacelike
hypersurface $\Sigma\subset\Mf$ which is intersected by any
nonextendable causal curve in \Mf\/ exactly once.  In other words,
$\Mf=D(\Sigma)$, where we have introduced the domain of causal
dependence $D(\Sigma)$ which is the set of all points $x\in\Mf$ such
that any causal curve through $x$ intersects $\Sigma$.  In the
globally hyperbolic case the manifold is topologically of the simple
form $\Mf=\R\times\Sigma$.  Many physically interesting spacetimes
like flat Minkowski spacetime, Schwarzschild spacetime as a black hole
model, or Robertson-Walker spacetimes as cosmological models fall into
this class.

In order to include more general models like anti-de Sitter spacetime,
this restriction can be somewhat weakened.  Crucial for the following
is mainly the existence of unique propagators as well as the existence
of a spinor bundle. Since our constructions are purely local, they
immediately carry over to all globally hyperbolic submanifolds of any
spacetime.

Globally hyperbolic spacetimes are time orientable, such that the
light cone, i.e.\ the set of all nonvanishing timelike covectors in
$T^\ast_x\Mf$, can be separated into a forward and backward light
cone, $V^+_x$ and $V^-_x$, continuously in $x$. The closed light cones
$\overline{V}^\pm_x$ include the lightlike covectors.  The time
orientation induces the separation of the causal future $J^+(x)$ and
past $J^-(x)$ of a point $x\in\Mf$ which are the sets of all points
that can be reached from $x$ by a future (past) directed causal curve.

While the tangent bundle \TM\/ and the cotangent bundle \TastM\/
naturally exist for any spacetime \Mf, the existence of spinors is
less trivial. In order to define spinors, the bundle of orthonormal
frames, which is a principal fibre bundle whose structure group is the
proper orthochronous Lorentz group, has to be lifted to a principal
fibre bundle with the universal covering group $\mathrm{SL}(2,\C)$ as
structure group, called the spin structure. The Dirac spinor bundle is
then defined as the associated $\C^4$-vector bundle.  It can be shown
that for any globally hyperbolic spacetime there exists such a spinor
bundle \DM\/ which in this case is just a trivial bundle over \Mf.
The construction of the spinor bundle leads naturally to a covariant
derivative on \DM\/ and the dual bundle \DastM\/ which will also be
denoted $\nabla$.  In local coordinates one has
$\nabla_\mu=\partial_\mu+\sigma_\mu$, where the $4\times 4$ matrices
$\sigma_\mu$ can be expressed in terms of the Christoffel symbols and
the Dirac matrices. For details see \cite{dimock}.

\subsection{The free Klein-Gordon field}

The simplest example of a quantum field theory is the free scalar
field, satisfying the Klein-Gordon equation
\begin{equation} 
  (\square_g+m^2)\phi(x)=0. 
\end{equation}
This is already the covariant field equation on a generic spacetime
$(\Mf,g)$ if $\square_g$ is the covariant wave operator
$\square_g=g^{\mu\nu}\nabla_\mu\nabla_\nu$ of the metric $g$. The
positive real constant $m$ plays the role of a mass.

Because of the global hyperbolicity of our spacetime, the Cauchy
problem for the Klein-Gordon operator has a unique solution, and there
exist unique retarded and advanced propagators, $\Delta_{ret}$ and
$\Delta_{adv}$, such that
\begin{equation} 
  (\square_g+m^2)\Delta_{ret,adv}=\Delta_{ret,adv}(\square_g+m^2)=\Eins 
\end{equation}
and $\supp(\Delta_{ret,adv} f)\subset J^\pm(\supp f)$.  We define the
commutator function $\Delta=\Delta_{ret}-\Delta_{adv}$ and identify
the operator $\Delta$ with the distribution
\begin{equation}
  \Delta:(f,g)\mapsto\Delta(f,g)=\int f(x)(\Delta g)(x)\mathrm{d}\mu(x),
\end{equation}
where $\mathrm{d}\mu$ is the volume element on \Mf.

In the quantum theory the field $\phi$ is represented as an operator
valued distribution such that we have smeared field operators
$\phi(f)$, $f\in\Cinfty_0(\Mf)$, acting on some Hilbert space with a
vacuum vector $\Omega$. They are required to satisfy the Klein-Gordon
equation and the canonical commutation relations
\begin{equation} 
  [\phi(f),\phi(g)]=i\Delta(f,g)\cdot\Eins.
\end{equation}

Since we investigate a free theory, we expect that all vacuum
expectation values of products of field operators are determined by
the two point function
\begin{equation}
  \Lambda(x,y)=<\Omega, \phi(x)\phi(y)\Omega>
\end{equation}
which is assumed to be a distribution $\Lambda\in\Dpr{\Mf\times\Mf}$.

Canonical quantization of the free Klein-Gordon field can be carried
through in close analogy to quantization on Minkowski spacetime (see
for example \cite{birell_davies}), and one ends up with a Fock space
representation of the field $\phi$ in terms of creation and
annihilation operators. However, the problem is that there is no
preferred mode expansion in terms of plain waves, because the Fourier
transform is not invariant under general coordinate transformations.
The crucial point is the ambiguous division into positive- and
negative-frequency modes.  Therefore this construction is highly
non-unique, and there is a large amount of unitarily inequivalent
representations, as was first observed by Fulling \cite{fulling}.  A
priori we do not know the `real' vacuum state $\Omega$, and as a
consequence there is no preferred interpretation of the theory in
terms of particles.  If the spacetime is asymptotically flat, one can
carry over the particle interpretation from the flat part to the whole
spacetime, but on a generic spacetime one has to face the fact that
quantum field theory is primarily a theory of fields, and no unique
particle interpretation can be expected.

From an axiomatic viewpoint, the vacuum state on Minkowski spacetime
is singled out by the requirement of its Poincar\'e invariance and the
spectrum condition which is the requirement that the energy-momentum
operator (the generator of translations) takes its spectrum in the
closed forward light cone.  In the absence of Poincar\'e invariance on
a generic spacetime these conditions no longer make any sense. While
one will still demand invariance under the symmetries of the spacetime
(if there are any), there is no simple analogue to the spectrum
condition since there are no generators of translations.  Thus one
central task of quantum field theory on curved spacetime is to
characterize physical states by finding a generalized spectrum
condition for generic spacetimes.

Usually the construction sketched above is performed in the algebraic
approach to quantum field theory. There the local net of observable
algebras can be defined in a unique way, but one still has to fix a
state on the algebra (which corresponds to the preparation of the
system). Again one restricts oneself to quasifree states that are
fixed once a two point distribution is specified, and via the GNS
construction the state induces a Hilbert space representation of the
observable algebra.  For details on this approach we refer to
\cite{wald_qft}.

\subsection{Hadamard states}

In order to single out a class of acceptable quantum states, one
physical requirement is that we should be able to formulate the
semi-classical Einstein equations in order to describe the back
reaction of the quantum field on the spacetime geometry:
\begin{equation} 
  R_{\mu\nu}-\frac{1}{2}g_{\mu\nu}R=8\pi\langle T_{\mu\nu}\rangle. 
\end{equation}
Here, $\langle T_{\mu\nu}\rangle$ is the expectation value of the
energy momentum tensor of the quantum field. Thus physical states
have to allow a unique definition of this expectation value.

It could be shown \cite{wald} that this requirement is fulfilled if we
demand that the two point function takes the following form which was
first given in \cite{dewitt_brehme} and since then intensively studied
(see \cite{fulling_buch} and the references therein):
\begin{equation}
  \Lambda(x,y)=\frac{1}{4\pi^2}\left(\frac{u(x,y)}{\sigma(x,y)}
             +v(x,y)\ln|\sigma(x,y)|+w(x,y)\right), 
\label{eq_hadamard}
\end{equation}
where $\sigma(x,y)$ denotes the quadratic geodesic distance, and $u$,
$v$, and $w$ are some smooth functions.

A state whose two point function is of this form is called a Hadamard
state.  The field equation and the commutation relations fix $u$ and
$v$, and the remaining freedom in the definition of a Hadamard state
lies in the choice of the smooth function $w$.

In order to properly define a distribution, one has to modify
(\ref{eq_hadamard}) by an $i\epsilon$-prescription in the denominator.
Also, since $\sigma$ is not defined globally, one must specify some
neighbourhood in which (\ref{eq_hadamard}) is supposed to hold, and it is
also understood that no further singularities exist for spacelike
separated points. To this end, one demands the convergence of a series
of distributions to $\Lambda$ in a causal normal neighbourhood (which
is basically a sufficiently small neighbourhood of a Cauchy surface).
Thereby the Hadamard condition becomes a global condition.  Although
the choice of a Cauchy surface enters into the definition, it can be
shown that the Hadamard condition is independent of this choice.  All
these details were worked out in \cite{kay_wald}, but most of them are
not essential for this work. What should be kept in mind is that the
vacuum state on Minkowski spacetime is Hadamard, that Cauchy evolution
respects the Hadamard singularity structure, and that two Hadamard
distributions differ only by a smooth part.

The main lesson of this rather sketchy definition is that the Hadamard
condition prescribes the singular short distance behaviour 
of the two point distribution, and it is
obvious that it would be very useful to have at hand some 
powerful mathematical methods to characterize this singularity structure.

\subsection{The free Dirac field}

We will now briefly outline the construction of the free Dirac field
on a curved spacetime. For details see \cite{dimock}.

Just like the Klein-Gordon equation, we can generalize the Dirac
equation to a spinor bundle over an arbitrary curved spacetime by
replacing the derivatives and the Dirac matrices by their covariant
counterparts. For a spinor field $\psi\in\Gamma(\Mf,\DM)$ and a
cospinor field $\overline{\psi}\in\Gamma(\Mf,\DastM)$ it reads
\begin{eqnarray}
  (-i{\nslash}+m)\psi = 0, &&
  \quad \nslash \psi = \gamma^\mu\nabla_\mu \psi; \\
  (i{\nslash}+m)\overline{\psi} = 0, &&
  \quad\nslash \overline{\psi} = (\nabla_\mu \overline{\psi})\gamma^\mu.
\end{eqnarray}
Here, $\gamma^\mu$ are the generalized Dirac matrices which satisfy
the anticommutation relations
\begin{equation}
\label{eq_gamma}
 \{\gamma^\mu,\gamma^\nu\}=2g^{\mu\nu}\cdot\Eins. 
\end{equation}
It can be shown that they form a covariant section
$\gamma\in\Gamma(\Mf,\TM\otimes\DM\otimes\DastM)$. Like in
(\ref{eq_gamma}), we will always choose a local frame and suppress the
spinor indices referring to $D_x\Mf\otimes D^*_x\Mf$ such that
$\gamma(x)$ is treated like a vector of $4\times4$ matrices.  We can
then define matrices $\sigma^{\mu\nu}, \gamma^5, \gamma^\mu\gamma^5$
and $\Eins$ in the usual way, which together with $\gamma^\mu$ give a
basis for $\DM\otimes\DastM$.  A further important property of
$\gamma$ is that it is covariantly constant: $\nabla\gamma=0$.  For
any covector field $\xi\in\Gamma(\Mf,\TastM)$ we define $\xislash$ as
contraction with $\gamma$, $\xislash\equiv\xi_\mu\gamma^\mu$.

For the quantization of the theory we again need propagators $S_{ret}$
and $S_{adv}$. For globally hyperbolic spacetimes it can be shown that
these propagators exist and are unique \cite{dimock}. Their difference
defines the anticommutator function $S=S_{ret}-S_{adv}$.

In the quantized theory the spinor field operator now has to be
smeared with a smooth cospinor field of compact support in order to
give an operator on the Hilbert space,
\begin{eqnarray}
  \psi(v),&&\quad v\in\Gamma_0(\Mf,\DastM);\nonumber\\
  \overline{\psi}(u),&&\quad u\in\Gamma_0(\Mf,\DM). 
\end{eqnarray}
The field operators are required to solve the Dirac equation and obey
the canonical anticommutation relations,
\begin{equation} 
  \{\psi(v),\overline{\psi}(u)\}=i S(v,u).
\end{equation}

Again, quasi-free states are completely characterized by their two
point distribution
\begin{equation}
  {\omega^+(x,y)}=<\Omega,\psi(x)\overline{\psi}(y)\Omega>
\end{equation}
which is now a vector-valued distribution
$\omega^+\in\Dpr{\Mf\times\Mf,\DM\boxtimes\DastM}$ taking values in
the bispinor bundle $\DM\boxtimes\DastM$. This denotes the outer
tensor product: the fibre over $(x,y)\in\Mf\times\Mf$ is
$D_x\Mf\otimes D^\ast_y\Mf$, and the first factor in the tensor
product transforms like a spinor in the point $x$, while the second
factor transforms like a cospinor in $y$.

In the quantization procedure on Minkowski spacetime one often makes
use of the fact that the squared Dirac operator gives back the
Klein-Gordon operator. On a curved spacetime this is expressed by
Lichnerowicz's identity:
\begin{equation}
  (-i{\nslash}+m)(i{\nslash}+m)\psi=(\square-\frac{1}{4}R+m^2)\psi.
\end{equation}
Here, $R$ is the curvature scalar, and
$\square=g^{\mu\nu}\nabla_\mu\nabla_\nu$ is the spinorial wave
operator acting on sections of \DM\@. Note that only the principal
part of $\square$ gives the scalar wave operator, and in general it
contains non-diagonal terms. Nevertheless, one obtains the propagators
$S_i$ for the Dirac equation by applying the Dirac operator to the
propagators $\Delta_i$ for this spinorial wave operator,
\begin{equation} 
  S_i=(i\nslash+m)\Delta_i.
\end{equation}
Analogously, one demands that the two point function $\omega^+$ can be
extracted from an auxiliary two point function $\tilde{\omega}$,
\begin{equation} 
  \omega^+(x,y)=\left(i\nslash_x+m\right)\tilde{\omega}(x,y). 
\end{equation}

Then $\tilde{\omega}$ is a solution of the spinorial Klein-Gordon
equation,
\begin{equation}  
   \left(\square_{x,y} -\frac{1}{4}R+m^2\right)\tilde{\omega}(x,y)=0, 
\end{equation}
and in order to define Hadamard states for the Dirac field one can
make the same ansatz as in the scalar case,
\begin{equation} 
\label{eq_had_dir}
   \tilde{\omega}=\frac{1}{4\pi^2}\left(\frac{\tilde{u}}{\sigma} 
   +\tilde{v}\ln|\sigma|+\tilde{w}\right),
\end{equation}
but now $\tilde{u}$, $\tilde{v}$, and $\tilde{w}$ are smooth
bispinor-valued functions. The rigorous definition was given by
K\"ohler \cite{koehler_diss}. Again, the singular part of the two point
function is fixed by geometry, and the freedom in the choice of a
Hadamard state lies in the choice of a smooth function $\tilde{w}$.


\section{Microlocal analysis}

Let us now introduce the mathematical tools we need to investigate the
singularities of distributions.  This theory of `microlocal analysis'
was developed by H\"ormander and Duistermaat in the seventies for
their analysis of partial differential equations, see their original
papers \cite{hoermander1,hoermander2} or the monographs
\cite{hoermander_buch,taylor}. For physical applications see for
example \cite{reed_simon,junker}.  It should be noted that similar
tools where independently developed by Bros and Iagolnitzer
\cite{iagolnitzer} in the context of analyticity properties of the
$S$-matrix.

The generalization to vector-valued distributions and the definition
of the polarization set was accomplished by Dencker in the eighties
\cite{dencker}, but it seems that is has not found its way into the
physical literature yet.

\subsection{Scalar distributions -- the wave front set}

As test function space we will take $\mathcal{D}(X)=\Cinfty_0(X)$, the
space of infinitely differentiable functions on an open subset
$X\subset\R^n$, equipped with the usual topology. Its dual is the
space of distributions, \Dpr{X}.

The roughest characterization of the singularities of a distribution
$u\in\Dpr{X}$ is the singular support, $\singsupp u$. It is the set of
all points $x\in X$ such that $u$ does not correspond to a smooth
function in any neighbourhood of $x$.

But this definition can be very much refined. The following statements
show that much information about the singularities is contained in the
Fourier transform:
\begin{enumerate}
\item If $u$ is of compact support, then there exists a Fourier
  transform $\widehat{u}$ which furthermore is a smooth function.
\item A distribution $u$ of compact support corresponds to a smooth
  function if and only if $\widehat{u}$ decays faster than any power,
  that is if for every $m\in\mathbb{N}$ there is a constant
  $C_m\in\R$ such that
  \begin{equation} 
     |\widehat{u}(\xi)|\leq C_m(1+|\xi|)^{-m} \quad \forall \xi\in\R^n. 
  \end{equation}
\end{enumerate}

Thus in order to investigate the singularities of a distribution
$u\in\Dpr{X}$ locally at the point $x\in X$, we are led to analyze the
Fourier transform $\widehat{\phi u}$ for test functions
$\phi\in\Cinfty_0(X)$ with $\phi(x)\neq 0$ which cut off the
singularities far away from $x$. Then $x\in\singsupp{u}$ iff there is
no $\phi$ such that $\widehat{\phi u}$ falls off rapidly in all
directions.

This immediately suggests two refinements of the singular support: we
may ask
\begin{itemize}
\item If $\widehat{\phi u}$ does not decay rapidly in all directions,
  which are those directions in Fourier space which are responsible
  for the singularity?
\item If $\widehat{\phi u}$ does not fall off rapidly, how fast does
  it decay?
\end{itemize}
The first question leads directly to the notion of the wave front set,
while the second question motivates the definition of local Sobolev
spaces $H^{s}(x)$. In combination one can define $H^{s}$-wave front
sets, but in the following we will only be interested in the
$\Cinfty$-case:

\begin{defi}
  Let $u\in\Dpr{X}$, $X\subset\R^n$. A point $(x,\xi)\in
  X\times(\R^n\setminus\{0\})$ is called a regular directed point of
  $u$ if there is a function $\phi\in\Cinfty_0(X)$, not vanishing at
  $x$, such that for any $m\in\mathbb{N}$ there is a constant
  $C_m\in\R$ with
  \begin{equation}   
     |\widehat{\phi u}(\xi')|\leq C_m(1+|\xi'|)^{-m}  
  \end{equation}
  for all $\xi'$ in a conic neighbourhood
  $\Gamma\subset\R^n\setminus\{0\}$ of $\xi$. (A neighbourhood
  $\Gamma$ is called conic if with $\xi\in\Gamma$ all points
  $t\cdot\xi$, $t\in\R^+$, are contained in $\Gamma$.)
  
  The wave front set of $u$, denoted \WF{u}, is the complement in
  $X\times(\R^n\setminus\{0\})$ of the set of regular directed points
  of $u$.
\end{defi}

\begin{example}
  The wave front set of the $\delta$-distribution
  $\delta\in\Dpr{\R^n}$ is
  \begin{equation} 
     \WF{\delta}=\bigl\{ (0,\xi)\,|\,\xi\in\R^n\setminus\{0\}\bigr\}. 
  \end{equation} 
  This can directly be seen from its Fourier transform
  \begin{equation}  
     \widehat{\phi\delta}(\xi)=\delta(e^{-i<\xi,\cdot>}\phi)=\phi(0)
  \end{equation}
  which does not decay rapidly in any direction.
\end{example}    

Our second example shows that the wave front set does not always
contain the whole Fourier space for all points of the singular
support:

\begin{example}
  The distribution $u\in\Dpr{\R}$, $u(x)=\frac{1}{x+i\epsilon},$ has
  got the wave front set
  \begin{equation} 
    \WF{u}=\bigl\{ (0,\xi)\,|\, \xi\in\R^+\setminus\{0\}\bigr\}
  \end{equation} 
  which can be seen from its Fourier transform
  \begin{equation}   
     \widehat{u}(\xi)=-i\sqrt{2\pi}\theta(\xi).
  \end{equation}
\end{example}    

We will now collect some basic properties of the wave front set
\cite{hoermander1}:
\begin{thm}
\label{thm_wf_basics}
\begin{enumerate}
\item \WF{u} is a closed conic subset of
  $X\times(\R^n\setminus\{0\})$.
\item The wave front set is a refinement of the singular support:
        \begin{equation} 
         \pi(\WF{u})=\singsupp{u}. 
        \end{equation}
        In particular, the wave front set of a smooth function is
        empty.
      \item For any partial differential (or more generally any
        pseudo-differential) operator $P$ we have the pseudolocal
        property
        \begin{equation}  
        \label{eq_pseudolocal}
           \WF{Pu}\subseteq\WF{u}.  
        \end{equation}
      \item Let $U,V\subset\R^n$, $u\in\Dpr{V}$, and
        $\chi:U\rightarrow V$ a diffeomorphism such that $\chi^*
        u\in\Dpr{U}$ is the distribution pulled back by $\chi$:
        $\chi^*u(f)=u(f\circ\chi^{-1})$. Then
        \begin{equation} 
          \WF{\chi^* u}=\chi^*\WF{u}\equiv
             \left\{\left(\chi^{-1}(x),\,^t\!\chi'(\chi^{-1}(x))\xi\right)\,|\,
             (x,\xi)\in\WF{u}  \right\}.
        \end{equation}
        In particular, under coordinate transformations the elements
        of the wave front set transform like covectors.  Thus the wave
        front set can be defined for distributions on differentiable
        manifolds by gluing together wave front sets over the
        coordinate patches, and for $u\in\Dpr{\Mf}$ we have
        $\WF{u}\subset\TastMon$.
\end{enumerate}
\end{thm}

The wave front set can be defined in a second way which will be
important for the investigation of solutions of partial differential
equations as well as for the generalization to vector-valued
distributions. But first we need some notions from the theory of
partial differential operators (PDOs):

A PDO $A$ is a polynomial of partial derivatives, $A=P(x,\partial)$. In
momentum space it acts like multiplication by the same polynomial of
momenta $P(x,i\xi)=\sigma_A(x,\xi)$ which is called the symbol of
$A$.  The leading order $a(x,\xi)$ of $\sigma_A$ is called the
principal symbol $\sigma_P(A)$.  The generalization to a manifold
\Mf\/ via local charts is obvious, and it should be noted that the
principal symbol is then a well-defined function on \TastM.

\begin{sloppypar}
Usually this is generalized to pseudo-differential operators
($\Psi$DOs), where $\sigma_A(x,\xi)$ may be a more general but still
well-behaved function, not necessarily a polynomial (for details see
\cite{taylor}). In this work we will only deal with classical
$\Psi$DOs whose symbol is an asymptotic sum of homogeneous terms.  The
set of classical $\Psi$DOs on $\Mf$ of order $m$ will be denoted
$L^m(\Mf)$.
\end{sloppypar}

The characteristic set of a $\Psi$DO $A\in L^m(\Mf)$ with principal
symbol $a=\sigma_P(A)$ is defined as
\begin{equation} 
  \ch(A)=\bigl\{(x,\xi)\in\TastMon\,|\,a(x,\xi)=0\bigr\} 
\end{equation}
and can be interpreted as the set of all directions $\xi$ suppressed
by $A$ to leading order, at a point $x$.

Now the following theorem makes precise the intuitive statement that
the singular directions of a distribution (which make up the wave
front set) are just those directions that have to be suppressed to
leading order by any operator that maps the distribution to a smooth
function.

\begin{thm}
\label{thm_wf_psdo}
For the wave front set of a distribution $u\in\Dpr{\Mf}$ we have
\begin{equation}  
  \WF{u}=\bigcap_{Au\in\Cinfty(\Mf)} \ch(A),
\end{equation}
where the intersection is taken over all pseudo-differential operators
$A\in L^0(\Mf)$ with $Au\in\Cinfty(\Mf)$.
\end{thm}

This property of the wave front set is particularly useful if we know
such an operator $A$ with $Au\in\Cinfty$, especially if $u$ is a
solution of a partial differential equation.  The first part of the
following theorem is then obvious, but for certain operators one can
make an even stronger statement which goes under the name `propagation
of singularities theorem' and was first proved by Duistermaat and
H\"ormander \cite{hoermander2}:
                           
\begin{thm}
\label{thm_prop_skal}
Let $P\in L^m(\Mf)$ be a $\Psi$DO on \Mf\/ with principal symbol $p$.
If $u\in\Dpr{\Mf}$ such that $Pu\in\Cinfty(\Mf)$, then
\begin{equation} 
  \WF{u}\subset \ch(P).
\end{equation}
If furthermore $P$ is of real principal type, then $\WF{u}$ is
invariant under the flow generated by the Hamiltonian vector field of
$p$.
\end{thm}

Here we have imposed a natural restriction on $P$ in order to obtain a
real Hamiltonian and a non-degenerate Hamiltonian flow:
\begin{defi}
  A pseudo-differential operator $P\in L^m(\Mf)$ is said to be of real
  principal type if its principal symbol $p(x,\xi)$ is real and for
  $p=0$ the Hamiltonian vector field $H_p$,
  \begin{equation} 
  H_p(x,\xi)=\sum_\mu\partial_{\xi_\mu}p(x,\xi)\partial_{x^\mu}
  -\sum_\mu\partial_{x^\mu}p(x,\xi)\partial_{\xi_\mu},
  \end{equation}
  does not vanish nor does it have the radial direction, that is
  $H_p\neq-\frac{\partial p}{\partial x^\mu}\frac{\partial}{\partial
    \xi_\mu}$.
\end{defi}

Thus the wave front set of a distribution $u$ with $Pu=0$ is made up
of integral curves of $H_p$ in $\ch(P)$ which are also called the null
bicharacteristics of $P$. Their projections onto $\Mf$ are called the
bicharacteristic curves of $P$ and constitute the singular support of
$u$.

To illustrate this theorem, let us take a look at a physical example,
the wave operator on a curved spacetime:

\begin{example}
  Let $(\Mf,g)$ be a spacetime, $f,b\in\Cinfty(\Mf)$, such that $0\neq
  f(x)\in\R\ \,\forall x\in\Mf$, and $a\in\Gamma(\Mf,\TM)$. The
  operator
\begin{equation}
   P=f\square+a^\mu\nabla_\mu+b,\quad\square=g^{\mu\nu}\nabla_\mu\nabla_\nu,
\end{equation}
has got the principal symbol
$p(x,\xi)=-f(x)g^{\mu\nu}(x)\xi_\mu\xi_\nu=-f\cdot\xi^2$. Hence, $P$ is
of real principal type.

For $(x,\xi)\in\ch(P)$ we see that the covector $\xi$ has to be
lightlike. Furthermore, for $\xi^2=0$ the function $p(x,\xi)=-\xi^2$
is a well-known Hamiltonian that generates the null geodesics which is
not affected by the nonvanishing factor $f$. Therefore the
bicharacteristic curves of $P$ are null geodesics $x(\tau)$.

From Hamilton's equations we obtain
\begin{equation}
  \frac{dx^\mu}{d\tau}=\frac{dp(x,\xi)}{d\xi_\mu}=-2f(x(\tau))g^{\mu\nu}\xi_\nu
\end{equation}
which means that $\xi$ is tangent to the geodesic. By
reparameterization of the geodesic one can achieve
$\xi(\tau)=\frac{dx}{d\tau}(\tau)$ for all values of the curve
parameter $\tau$. The factor, however, is irrelevant in what follows
since the wave front set is conic in $\xi$.

Thus the null bicharacteristics of $P$ are curves
$(x,\xi)(\tau)\in\TastMon$ such that $x(\tau)$ describes a lightlike
geodesic and $\xi(\tau)$ is tangent to $x(\tau)$ for all values of
$\tau$.

Applying theorem~\ref{thm_prop_skal}, we conclude that the singular
support of a distribution $u$ which solves the wave equation, i.e.\
$Pu=0$, is a union of null geodesics, while the wave front set
additionally indicates the covectors tangent to the geodesics in the
singular support. Note that the geodesics are directed and that the
wave front set may contain only the future or past directed covectors.

If we take light rays as distributional solutions of the wave
equation, this result can be interpreted as the well-known fact that
light propagates along lightlike geodesics.
\end{example}

\subsection{Vector-valued distributions -- the polarization set}

We will now go over to vector-valued distributions
$u\in\Dpr{\Mf,\CN}$, i.e.\ vectors $u=(u_i)_{i=1\dots N}$ of
distributions $u_i\in\Dpr{\Mf}$. This section is mainly a review of
\cite{dencker}.

The wave front set of a vector-valued distribution
$u=(u_i)\in\Dpr{\Mf,\CN}$ is just defined as the union of the wave
front sets of all its components:
\begin{equation}  
  \WF{u}=\bigcup_{i=1}^N \WF{u_i}.   
\end{equation}
Because of theorem~\ref{thm_wf_psdo}, this is nothing but
\begin{equation} 
  \WF{u}=\bigcup_{i=1}^N\bigcap_{Au_i\in\Cinfty(\Mf)} \ch(A).
\end{equation}

The wave front set does not contain any information about the
components of the distribution that are singular. In order to specify
the singular directions in the vector space $\CN$, one could consider
vector-valued operators that map the vector-valued distribution to a
smooth scalar function, instead of just looking at scalar operators
mapping the individual components to smooth functions.  This approach
leads to the definition of the polarization set:

\begin{defi}
  The polarization set of a distribution $u\in\Dpr{\Mf,\CN}$ is
  defined as
\begin{equation} 
  \WFpol{u}=\bigcap_{Au\in\Cinfty(\Mf)} \mathcal{N}_A,
\end{equation}
\begin{equation} 
  \mathcal{N}_A=\bigl\{(x,\xi; w)\in (\TastMon)\times\CN\,|\,w\in \ker
  a(x,\xi)\bigr\}, 
\end{equation}
where the intersection is taken over all $1\times N$ systems $A\in
L^0(\Mf)^N$ of classical $\Psi$DOs with principal symbol $a$, and
$\ker a(x,\xi)$ is the kernel of the matrix $a(x,\xi)$.
\end{defi}

Obviously, $(\TastMon)\times \{0\}\subset\WFpol{u}$ for any
$u\in\Dpr{\Mf,\CN}$. Furthermore, the polarization set is closed,
linear in the fibre and conic in the $\xi$-variable.

For scalar distributions ($N=1$) the polarization set contains the
same information as the wave front set. For arbitrary $N$, we get back
the wave front set by projecting the nontrivial points onto the
cotangent bundle:

\begin{thm}
  Let $u\in\Dpr{\Mf,\CN}$ and
  $\pi_{1,2}:\TastM\times\CN\rightarrow\TastM$ be the projection onto
  the cotangent bundle: $\pi_{1,2}(x,\xi;w)=(x,\xi)$. Then
\begin{equation} 
  \pi_{1,2}(\WFpol{u}\setminus (\TastM\times \{0\}))=\WF{u}. 
\end{equation}
\end{thm}

In this way the polarization set is a refinement of the wave front
set. In addition, it contains information about the directions in the
additional vector space in which the distribution is singular.

\begin{example}
  Let $u=(u_1,u_2)\in\Dpr{\Mf,\C^2}$ and $(y,\eta)\notin \WF{u_1}$.
  Then
  \begin{equation} 
     \WFpol{u}\subseteq\bigl\{(x,\xi;(0,z))\in (\TastMon)\times\C^2,\,
     z\in\C\bigr\}  
  \end{equation}
  over a conic neighbourhood of $(y,\eta)$.
\end{example}

The polarization set indicates only the most singular directions, even
if the projection of the distribution on other directions is also
singular, as the following example shows:

\begin{example}
  Let $u=(v,\Delta v)\in\Dpr{\R^n,\C^2}$, where $v\in\Dpr{\R^n}$ and
  $\Delta$ is the Laplacian on $\R^n$. Then
  \begin{equation} 
     \WFpol{u}\subseteq\bigl\{(x,\xi;(0,z))\in
     (T^\ast\R^n\setminus\mathbf{0})\times\C^2,\,z\in\C\bigr\},  
  \end{equation}
  because $\Delta u_1-u_2=0$ and $\ker
  \sigma_P(\Delta,-\Eins)(x,\xi)=\ker (-\xi^2,0)=\{(0,z),\,z\in \C\}$.
\end{example}

Our third example shows that the direction of the strongest
singularities in $\C^N$ can depend on the singular direction in the
cotangent bundle:

\begin{example}
  Let $u=\nabla\delta^{(2)}=\left(\partial_{x_1}\delta^{(2)},
    \partial_{x_2}\delta^{(2)}\right)\in\Dpr{\R^2,\C^2}$.  Then
  \begin{equation} 
    \WFpol{u}\subseteq\bigl\{(0,\xi;\lambda\cdot\xi)\in
    (T^\ast\R^2\setminus\mathbf{0})\times\C^2,\,\lambda\in\C\bigr\},  
  \end{equation}
  since $u$ solves the equation $(-\partial_{x_2},\partial_{x_1})\cdot
  u=0$, and $\ker (-i\xi_2,i\xi_1)=\C\cdot(\xi_1,\xi_2)$ for $\xi\neq
  0$.
\end{example}

The polarization set has got an interesting transformation property
under the action of a system of partial differential operators:

\begin{thm}
\label{thm_wfpol_trans}
Let $A$ be an $M\times N$ system of pseudo-differential operators on
\Mf\/ with principal symbol $a(x,\xi)$, and $u\in\Dpr{\Mf,\CN}$. Then
\begin{equation} 
\label{eq_wfpoltrans1} 
  a(\WFpol{u})\subseteq \WFpol{Au}, 
\end{equation}
where a acts on the fibre: $a(x,\xi;w)=(x,\xi;a(x,\xi)w)$.

If $E$ is an $N\times N$ system of pseudo-differential operators on
\Mf\/ and its principal symbol $e(x,\xi)$ is not characteristic at
$(y,\eta)\in\TastMon$, i.e.\ $\det e(y,\eta)\neq0$, then
\begin{equation} 
\label{eq_wfpoltrans2}
  e(\WFpol{u})=\WFpol{Eu} 
\end{equation}
over a conic neighbourhood of $(y,\eta)$.
\end{thm}

Note the different order of the inclusion in (\ref{eq_wfpoltrans1})
compared to (\ref{eq_pseudolocal}). Since the polarization set only
indicates the most singular directions, the projection on orthogonal
directions can change the polarization set substantially.

From (\ref{eq_wfpoltrans2}) we learn that the polarization set behaves
covariantly under a change of basis in the vector space $\C^N$. Thus
the polarization set of a distribution $u\in\Dpr{\Mf,E}$ taking values
in a vector bundle $E\rightarrow\Mf$ can be defined by gluing together
the polarization sets of local trivializations. This gives a
well-defined subset $\WFpol{u}\subset \pi^\ast E$ of the vector
bundle $E$ lifted over the cotangent bundle.

Like in the scalar case, we are interested in the propagation of
singularities for a solution of a system of partial differential
equations.  Again it turns out that there is a powerful theorem for a
restricted class of operators.

\begin{defi}
  An $N\times N$ system $P$ of pseudo-differential operators on \Mf\/
  with principal symbol $p(x,\xi)$ is said to be of real principal
  type at $(y,\eta)\in\TastMon$ if there is an $N\times N$ symbol
  $\tilde{p}(x,\xi)$ such that
  \begin{equation} 
    \tilde{p}(x,\xi) p(x,\xi)=q(x,\xi)\cdot \Eins 
  \end{equation}
  in a neighbourhood of $(y,\eta)$, with a scalar symbol $q(x,\xi)$ of
  real principal type.  We say that $P$ is of real principal type in
  $\mathcal{O}\subseteq\TastMon$ if it is at all points
  $(y,\eta)\in\mathcal{O}$.
\end{defi}

The choice of $\tilde{p}$ and $q$ is not unique, but it can be shown
that the only freedom is the multiplication by a smooth non-vanishing
function on \TastMon.

From now on let $P$ be an $N\times N$ system of classical
pseudo-differential operators $P$ of real principal type of order $m$,
$u$ a solution of $Pu=0$, and locally $\tilde{p}(x,\xi)$ and
$q(x,\xi)$ are chosen as above such that
\begin{equation} 
\label{eq_rpt}
  \tilde{p} p=q\cdot\Eins. 
\end{equation}

We define the set
\begin{equation}  
  \Omega_P=\bigl\{(x,\xi)\in\TastMon\,|\,\det p(x,\xi)=0\bigr\}.  
\end{equation}
Because of (\ref{eq_rpt}), we have $\Omega_P=q^{-1}(0)$ locally.  If
$Pu\in\Cinfty$, for the existence of nontrivial elements
$(x,\xi;w)\in\mathcal{N}_P$ in the polarization set over a point
$(x,\xi)$, by definition it is necessary that $w\in\ker p(x,\xi)$, that
is, $\ker p(x,\xi)\neq\{0\}$. This means that $(x,\xi)$ and thus the
whole set $\WF{u}$ has to be a subset of $\Omega_P$.

It can now easily be shown that the wave front set is a union of null
bicharacteristics of $q$ in $\Omega_P$.  Though the bicharacteristics
are not independent of the choice of $q$, we will just call them the
bicharacteristics in $\Omega_P$ without referring to a special choice
of $q$. A different choice of $q$ only leads to a multiplication of
the $\xi$-variable by a scalar factor which is irrelevant since the
wave front set is conic in $\xi$.  In fact, in general it is not even
possible to choose $q$ globally, and several coordinate patches must
be glued together anyhow.

Once we know the wave front set of the distribution $u$, the remaining
task is to calculate the polarization vectors over the points in the
wave front set. It turns out that these vectors follow a simple
parallel transport law along the bicharacteristics that form the wave
front set.

This parallel transport, which we will introduce in a second, does not
only depend on the principal symbol as in the scalar case, but also
the subprincipal symbol contributes. The symbol of $P$ is a sum of
homogeneous terms,
\begin{equation}  
  \sigma(P)(x,\xi)=p(x,\xi)+p_{m-1}(x,\xi)+p_{m-2}(x,\xi)+\dots, 
\end{equation}
where $p=\sigma_P(P)$ is the principal symbol, and $p_j$ is
homogeneous of order $j$.  The subprincipal symbol is now defined as
\begin{equation}  
  p^s(x,\xi)=p_{m-1}(x,\xi)-\frac{1}{2i}\sum_\mu\frac{\partial^2
  p(x,\xi)}{\partial x^\mu \partial \xi_\mu}. 
\end{equation}

Furthermore let
\begin{equation} 
  \{\tilde{p},p\}= \sum_\mu\frac{\partial
  \tilde{p}(x,\xi)}{\partial\xi_\mu}\frac{\partial p(x,\xi)}{\partial x^\mu}
  -\sum_\mu\frac{\partial\tilde{p}(x,\xi)}{\partial{x^\mu}}\frac{\partial
  p(x,\xi)}{\partial{\xi_\mu}}
\end{equation}
be the Poisson bracket and 
\begin{equation}
    H_q(x,\xi)=\sum_\mu \left(\frac{\partial
    q(x,\xi)}{\partial\xi_\mu}\frac{\partial}{\partial x^\mu} -
    \frac{\partial q(x,\xi)}{\partial x^\mu}\frac{\partial}{\partial
    \xi_\mu}\right)
\end{equation}
the Hamiltonian vector field of $q$.

\begin{defi}
  For a smooth section $w$ of the vector bundle $(\TastMon)\times\CN$,
  we define Dencker's connection as
\begin{equation}
\label{eq_dencker}
  D_P w=H_q w +\frac{1}{2}\{\tilde{p},p\} w + i\tilde{p}p^s w.
\end{equation}
\end{defi}

This is a partial connection along the bicharacteristics in
$\Omega_P$, that is a connection on all $\CN$-vector bundles over
these bicharacteristics.  Furthermore, it can be shown that
(\ref{eq_dencker}) defines a partial connection in $\mathcal{N}_P$,
this means that for every vector field $w$ along a bicharacteristic
$\gamma$ in $\Omega_P$ we have $D_Pw\in\ker p$ along $\gamma$ if and
only if $w\in\ker p$ along $\gamma$. The equation $D_Pw=0$ can then be
solved with $(x,\xi;w)\in\mathcal{N}_P$, which is a necessary
condition for elements of the polarization set.

Again, the definition of the connection depends on the choice of
$\tilde{p}$ and $q$, but it can be shown that a different choice
changes a solution of the equation $D_Pw=0$ in $\mathcal{N}_P$ only by
a scalar factor.  Therefore, we define a Hamilton orbit of a system
$P$ of real principal type as a line bundle
$L\subseteq\mathcal{N}_P|_\gamma$ over a bicharacteristic $\gamma$ in
$\Omega_P$ which is spanned by a section $w$ satisfying the equation
$D_Pw=0$, i.e., that is parallel with respect to Dencker's connection.
These Hamilton orbits are then independent of the choice of $\tilde{p}$.

Now we are prepared to state the main result of \cite{dencker}, the
theorem on the propagation of singularities for vector-valued
distributions:

\begin{thm}
\label{thm_prop_vect}
Let $P$ be an $N\times N$ system of classical pseudo-differential
operators over a manifold \Mf, and $u\in\Dpr{\Mf,\CN}$. Furthermore,
let $P$ be of real principal type at $(y,\eta)\in\Omega_P$, and
$(y,\eta)\notin\WF{Pu}$. Then, over a neighbourhood of $(y,\eta)$ in
$\Omega_P$, $\WFpol{u}$ is a union of Hamilton orbits of $P$.
\end{thm}

Though it is not obvious from the definition (\ref{eq_dencker}), in
interesting cases Dencker's connection takes on a very simple form. We
will illustrate this by a physical example that was already discussed
by Radzikowski \cite{radzikowski_unp} in a slightly modified way.

\begin{example}
  We investigate Maxwell's equations on a spacetime $(\Mf,g)$. For a
  vector field $A\in\Gamma(\Mf,\TM)$ in Lorentz gauge ($\nabla_\mu
  A^\mu=0$), in a local coordinate frame they read
\begin{equation} 
  \square_g A^\nu-{R^\nu}_\mu A^\mu=0,\quad
  \square_g=g^{\rho\sigma}\nabla_\rho\nabla_\sigma.  
\end{equation}
We want to calculate Dencker's connection for the operator
\begin{equation} 
  {P^\nu}_\mu={{\square_g}^\nu}_\mu-{R^\nu}_\mu=
  g^{\rho\sigma}(\delta^\nu_\mu \partial_\rho\partial_\sigma
  +{\Gamma^\nu}_{\rho\mu}\partial_\sigma
  -\delta^\nu_\mu{\Gamma^\lambda}_{\rho\sigma}\partial_\lambda
  +{\Gamma^\nu}_{\sigma\mu}\partial_\rho)+\dots, 
\end{equation}
where the dots stand for lower order terms that do not contribute to
Dencker's connection.

The principal symbol of $P$ is
\begin{equation}
  {p(x,\xi)^\nu}_\mu=-g^{\rho\sigma}(x)\xi_\rho\xi_\sigma
  \delta^\nu_\mu,
\end{equation}
so that we can choose
\begin{equation} 
  {{{\tilde{p}^\lambda}}\/}_\nu=\sqrt{-g}\delta^\lambda_\nu,\qquad
  q(x,\xi)=-\sqrt{-g}g^{\rho\sigma}\xi_\rho\xi_\sigma
\end{equation}
to get $\tilde{p}p=q\cdot\Eins$.  The factor $\sqrt{-g}$ is introduced
for no obvious reason, but it turns out that it simplifies the
calculation.  We see that $q$ is a scalar symbol of real principal
type, therefore $P$ is of real principal type.

The bicharacteristic curves of $q$ are again the null geodesics, and
the space $\Omega_P=q^{-1}(0)$ consists of all lightlike vectors in
\TastMon.  Since $p\propto \Eins$, the space $\ker p(x,\xi)$ on which
Dencker's connection acts is the whole tangent space.

Putting together the different terms in (\ref{eq_dencker}) and using
well-known identities for the Christoffel symbols and Hamilton's
equations, we arrive at
\begin{eqnarray}
\nonumber
(D_P w)^\nu
&=&\left(H_q w +\frac{1}{2}\{\tilde{p},p\} w + i\tilde{p}p^s w\right)^\nu\\
\nonumber
&=&\frac{dw^\nu}{d\tau}+{\Gamma^\nu}_{\rho\mu}\dot{x}^\rho
  w^\mu\\
&=&\left((\pi^*\nabla_\tau) w\right)^\nu,
\end{eqnarray}
for a vector field $w(x(\tau),\xi(\tau))$ along the null
bicharacteristics.  That is, Dencker's connection for $P$ (with
respect to our choice of $\tilde{p}$) is just the usual Levi-Civit\'a
connection, lifted over the cotangent bundle and restricted to the
bicharacteristics. Thus a vector field over one of the
bicharacteristics is parallel with respect to Dencker's connection
(and thereby generates a Hamilton orbit) iff the projected vector
field over the characteristic curve is parallel with respect to the
Levi-Civit\'a connection.

Thus, following the theorem on the propagation of singularities, the
polarization set of a solution of Maxwell's equations is a union of
such Hamilton orbits, that is it consists of curves $(x,\xi;w)(\tau)
\in\pi^\ast\TM$ such that
\begin{itemize}
\item
$x(\tau)$ describes a null geodesic,
\item
$\xi(\tau)$ is tangent to the geodesic and
\item
$w(\tau)$ is parallel transported along the geodesic.
\end{itemize}

In physical language, we have reproduced the well-known result on the
propagation of light in a curved spacetime: light travels along
lightlike geodesics, while the polarization vector is parallel
transported along the path.

In particular this means that spacetime curvature cannot lead to
double refraction which is only possible if matter effects are
introduced into Maxwell's equations. Double refraction may then occur
at points where the operator $P$ is no longer of real principal type.
This was also investigated by Dencker \cite{dencker2}.
 \end{example}


\section{Singularity structure of the two point function}

We are now going to investigate the singularity structure of the two
point function of Hadamard states using the microlocal techniques that
were introduced in the previous section.

The wave front set of a scalar Hadamard distribution was first
calculated by Radzikowski \cite{radzikowski}. After we have reviewed
his results, we will derive an analogous result for the wave front set
of the two point function of the free Dirac field and furthermore
calculate its polarization set.

\subsection{Klein-Gordon field}

Let us first state the main result by Radzikowski \cite{radzikowski}:

\begin{thm}
\label{thm_kg}
A quasi-free state of the Klein-Gordon field on a globally hyperbolic
spacetime \Mf\/ is a Hadamard state if and only if its two point
distribution $\Lambda$ has got the following wave front set:
  \begin{equation}
  \label{eq_wf_kg}
    \WF{\Lambda}=\bigl\{(x,y;\xi,-\eta)\in 
    \TastMMon\,|\,(x,\xi)\sim (y,\eta),\xi\in \overline{V}_x^+\bigr\}.  
  \end{equation}
  The equivalence relation $(x,\xi)\sim(y,\eta)$ means that there is a
  lightlike geodesic $\gamma$ connecting $x$ and $y$, such that at the
  point $x$ the covector $\xi$ is tangent to $\gamma$ and $\eta$ is
  the vector parallel transported along the curve $\gamma$ at $y$
  which is again tangent to $\gamma$.  On the diagonal
  $(x,\xi)\sim(x,\eta)$ if $\xi$ is lightlike and $\xi=\eta$.
\end{thm}

Thus the two point distribution of a Hadamard state is singular only
for pairs of points that can be connected by a lightlike geodesic. In
addition, the condition $\xi\in \overline{V}_x^+$ expresses the fact
that only positive frequencies contribute and can be viewed as the
microlocal remnant of the spectrum condition. The obvious advantage of
our formalism from microlocal analysis is that the spectrum condition
(for the free scalar field) is now formulated also for curved
spacetimes.

This microlocal characterization of the Hadamard condition opens the
door for further investigations of Hadamard states using the powerful
tools from microlocal analysis. As a first example, in \cite{junker}
it was shown that certain states that were known long before, the
ground- and KMS-states on static spacetimes and adiabatic states of
infinite order on Robertson-Walker spacetimes, are indeed Hadamard
states.

We will now briefly review K\"ohler's proof \cite{koehler_diss} that
the two point function of a Hadamard state has got the given wave
front set, since it will lead us in the generalization to the Dirac
field.

First we take a look at the two point function on flat spacetime:

\begin{thm}
\label{thm_wf_mink}
On Minkowski spacetime the two point function of the free Klein-Gordon
field in the vacuum state has got the wave front set
\begin{eqnarray}
  \WF{\Lambda}&=&\bigl\{ (x,y;\xi,-\xi)\in
  T^\ast(\R^{1,3}\times\R^{1,3})\,|
\protect{\begin{array}[t]{l}x\neq y, (x-y)^2=0,\bigr. \nonumber\\
         \bigl.\xi||(x-y), \xi_0>0\bigr\}\end{array}} \\
  &&\cup\ \bigl\{ (x,x;\xi,-\xi)\in
    T^\ast(\R^{1,3}\times\R^{1,3})\,|\,\xi^2=0, \xi_0>0\bigr\} ,
  \end{eqnarray}
  i.e.\ it is of the form (\ref{eq_wf_kg}).
\end{thm}

A complete proof can be found in \cite{reed_simon}, or the wave front
set can be read off from the well-known Fourier transform
\begin{equation}
\widehat{\Lambda}(\xi,\eta)=
(2\pi)^{-1}\delta(\xi+\eta)\theta(\xi_0)\delta(\xi^2-m^2).
\end{equation}

The extension of this theorem to arbitrary spacetimes is done by
deforming the spacetime such that the metric is flat in one part of
the spacetime. The known singularity structure in the flat part can
then be shifted to the curved part by applying H\"ormander's theorem
on the propagation of singularities.

The proof makes use of the following theorem which was proved in
\cite{koehler_diss} based on ideas from \cite{fnw}:

\begin{thm}
  Let $(\Mf,g)$ be a globally hyperbolic spacetime and $x$ a point on
  a Cauchy surface $\Sigma\subset\Mf$.  Then there is a neighbourhood
  $U$ of $x$ and a globally hyperbolic spacetime
  $(\tilde{\Mf},\tilde{g})$ with the following properties:
\begin{enumerate}
\item A causal normal neighbourhood $\mathcal{N}$ of $\Sigma$ with
  $U\subset\mathcal{N}$ is isometric to a neighbourhood
  $\tilde{\mathcal{N}}$ in $\tilde{\Mf}$, and the isometry
  $\rho:\mathcal{N}\rightarrow \tilde{\mathcal{N}}$ maps $\Sigma$ to a
  Cauchy surface $\tilde{\Sigma}\subset\tilde{\mathcal{N}}$.
\item There is a spacelike hypersurface $\hat{\Sigma}$ together with a
  neighbourhood $\hat{U}$ in $\tilde{\Mf}$ such that $\tilde{g}$,
  restricted to $\hat{U}$, is the Minkowski metric and
  $\rho(U)\equiv\tilde{U}\subset D(\hat{\Sigma})$.
\end{enumerate}
\end{thm}

The two point function $\Lambda$ which is given on the neighbourhood
$\mathcal{N}$ of the Cauchy surface $\Sigma$ and thereby also on
$\tilde{\mathcal{N}}=\rho(\mathcal{N})$ now induces a Hadamard two
point distribution $\tilde{\Lambda}$ on the whole deformed spacetime
$\tilde{\Mf}$ such that $\Lambda|_{\mathcal{N}\times\mathcal{N}}=
\rho^\ast(\tilde{\Lambda}|_{\tilde{\mathcal{N}}\times\tilde{\mathcal{N}}})$.

In the flat part of $\tilde{\Mf}$ we already know the wave front set
of the distribution $\tilde{\Lambda}$. Since only local properties of
the spacetime in a neighbourhood of a Cauchy surface enter into the
Hadamard condition and the metric on $\hat{U}$ is flat, on
$\hat{U}\times\hat{U}$ we have because of theorem~\ref{thm_wf_mink}:
\begin{equation} 
  (\hat{x},\hat{y};\hat{\xi},-\hat{\eta})\in 
  \WF{\tilde{\Lambda}|_{\hat{U}\times\hat{U}}}\ \Leftrightarrow\ 
  (\hat{x},\hat{\xi})\sim (\hat{y},\hat{\eta}),\hat{\xi}\in
  \overline{V}_{\hat{x}}^+. 
\end{equation}

$\tilde{\Lambda}$ solves the Klein-Gordon equation in the second
variable, therefore we can apply theorem~\ref{thm_prop_skal} on the
propagation of singularities for the operator
$\Eins\otimes(\square+m^2)$ to get the singular directions of
$\tilde{\Lambda}$ for points in $\hat{U}\times\tilde{U}$. From our
investigation of the wave operator in the previous section, we learned
that two points $(x,\xi)$ and $(y,\eta)$ lie on the same null
bicharacteristic for the operator $\square+m^2$ iff
$(x,\xi)\sim(y,\eta)$.  Because of $\tilde{U}\subset D(\hat{U})$,
every null geodesic through $\tilde{U}$ intersects $\hat{U}$, and
theorem~\ref{thm_prop_skal} tells us that
\begin{eqnarray}
  (\hat{x},y;\hat{\xi},-\eta)\in
  \WF{\tilde{\Lambda}|_{\hat{U}\times\tilde{U}}}
  &\Leftrightarrow&(\hat{x},\hat{y};\hat{\xi},-\hat{\eta})\in
  \WF{\tilde{\Lambda}|_{\hat{U}\times\hat{U}}},
  (y,\eta)\sim (\hat{y},\hat{\eta})\nonumber \\
  &\Leftrightarrow&(y,\eta)\sim(\hat{x},\hat{\xi}), \hat{\xi}\in
  \overline{V}_{\hat{x}}^+.
\end{eqnarray}

The same argument for the operator $(\square+m^2)\otimes \Eins$ gives
us the wave front set for points in $\tilde{U}\times\tilde{U}$:
\begin{equation} 
  (x,y;\xi,-\eta)\in\WF{\tilde{\Lambda}|_{\tilde{U}\times\tilde{U}}}
  \ \Leftrightarrow\ (x,\xi)\sim (\hat{x},\hat{\xi}),
  (y,\eta)\sim(\hat{x},\hat{\xi}), \hat{\xi}\in
  \overline{V}_{\hat{x}}^+, 
\end{equation}
and therefore
\begin{equation} 
  \WF{\tilde{\Lambda}|_{\tilde{U}\times\tilde{U}}}=
  \bigl\{(x,y;\xi,-\eta)\in T^\ast(\tilde{U}\times\tilde{U})\setminus \mathbf{0} 
  \,|\, (x,\xi)\sim (y,\eta),\xi\in \overline{V}_x^+\bigr\}.  
\end{equation}

Now $\Lambda$ is the distribution pulled back by the isometry $\rho$,
and because of theorem~\ref{thm_wf_basics}.4 also the wave front set
of $\Lambda$ in $U\times U$ is of this form.

Using the same line of arguments again in the undeformed spacetime
$\Mf$, we finally obtain the wave front set of $\Lambda$ for arbitrary
pairs of points $(x,y)\in \Mf\times\Mf$:
\begin{equation}  
  \WF{\Lambda}=\bigl\{(x,y;\xi,-\eta)\in\TastMMon\,|\,
  (x,\xi)\sim (y,\eta),\xi\in \overline{V}_x^+\bigl\},
\end{equation}
and we have finished the proof.  \qed

\subsection{Dirac field}

Let us now turn to Hadamard states of the free Dirac field. We first
state our main result:

\begin{thm}
\label{thm_dirac}
Let $\omega$ be a Hadamard state of the free Dirac field on a globally
hyperbolic spacetime \Mf.  Then its two point function $\omega^+$ has
got the following wave front and polarization sets:
\begin{eqnarray}
\label{eq_wf_dirac}
      \WF{\omega^+}&=&\bigl\{(x,y;\xi,-\eta)\in
    \TastMMon\,|\,(x,\xi)\sim (y,\eta),\xi\in \overline{V}^+_x\bigr\} ,
\end{eqnarray}
\begin{eqnarray}
 \WFpol{\omega^+}&=&\bigl\{(x,y;\xi,\eta;w)\in\pi^\ast(\DM\boxtimes\DastM)\,|\,
\bigr. \nonumber  \\ 
\label{wfpol_dirac}
    &&\bigl.(x,y;\xi,\eta)\in\WF{\omega^+}; 
    (\Eins\otimes\mathcal{J}_\gamma(x,y))w=\lambda\cdot\slash{\xi},
    \lambda\in\C \bigr\}.
\end{eqnarray}
Here, $\mathcal{J}_\gamma(x,y):D^*_y\Mf\rightarrow D_x^*\Mf$ denotes
the parallel transport in \DastM\/ along the geodesic $\gamma$
connecting $x$ and $y$, such that $\xi$ is tangent to $\gamma$ in the
point $x$.
\end{thm}

The proof is similar to that for the scalar case: We will first
calculate the polarization set on Minkowski spacetime and afterwards
shift the result to our spacetime \Mf\/ by making use of Dencker's
theorem on the propagation of singularities.

\begin{thm}
  The two point function $\omega^+$ of a Hadamard state of the free
  Dirac field on Minkowski spacetime $\Mf=\R^{1,3}$ is (up to a smooth
  part) of the form
\begin{equation}  
   \omega^+=\left[ (i\slash{\partial}+m)\otimes\Eins\right]
   (\Lambda\cdot\Eins),  
\end{equation}
where $\Lambda$ is the two point function of a Hadamard state of the
free Klein-Gordon field.  Its polarization set is
\begin{eqnarray}
  \lefteqn{ \WFpol{\omega^+}=} \nonumber \\
  &&\bigl\{(x,y;\xi,\eta;\lambda\cdot\!\slash{\xi})\in\pi^\ast(\DM\boxtimes\DastM)
  \,|\,(x,y;\xi,\eta)\in\WF{\Lambda}, \lambda\in\C \bigr\}. 
\end{eqnarray}
\end{thm}

\begin{proof}
  The bundle $\DM\boxtimes\DastM$ over Minkowski spacetime
  $\Mf=\R^{1,3}$ is the trivial bundle
  $(\Mf\times\Mf)\times\C^{4\times 4}$. Thus we can simplify our
  notation by identifying the spinor spaces over all points. Also, the
  covariant derivative coincides with the partial derivative of the
  individual components, and the curvature scalar vanishes.  Therefore
  the functions $\tilde{u}$ and $\tilde{v}$ in the definition
  (\ref{eq_had_dir}) of a Hadamard distribution have the form
  $\tilde{u}=u\cdot\Eins$, $\tilde{v}=v\cdot\Eins$ with the
  corresponding scalar functions $u$ and $v$.  That is, the auxiliary
  two point function of any Hadamard state of the free Dirac field on
  Minkowski spacetime is (up to a smooth part) a multiple of the unit
  matrix, and the nonvanishing components are two point functions of
  scalar Hadamard states.  Since these are fixed up to a smooth part,
  the auxiliary two point function is of the form
  $\tilde{\omega}=\Lambda\cdot\Eins$ with a scalar Hadamard
  distribution $\Lambda$ (up to a smooth part), and we have
\begin{equation}
  \label{eq_omegaplus}
  \omega^+=[(i\slash{\partial}+m)\otimes\Eins](\Lambda\cdot\Eins).
\end{equation}

From theorem~\ref{thm_kg} we know that the wave front set of
$\Lambda\cdot\Eins$ has the form (\ref{eq_wf_kg}), and its
polarization set is obviously
\begin{equation} 
   \WFpol{\Lambda\cdot\Eins}=\bigl\{(x,y;\xi,\eta;\lambda\cdot\Eins)
   \,|\,(x,y;\xi,\eta)\in\WF{\Lambda}, \lambda\in\C\bigr\}. 
\end{equation}

We obtain $\omega^+$ from $\Lambda\cdot\Eins$ by application of the
operator
\begin{equation}
   A=[(i\slash{\partial}+m)\otimes\Eins]
\end{equation}
with principal symbol $a(x,y;\xi,\eta)=-\xislash\otimes\Eins$.
Following theorem~\ref{thm_wf_basics}, this does not enlarge the wave
front set.  In addition, from theorem~\ref{thm_wfpol_trans} we have
the following restriction on the polarization set of
$\omega^+=A(\Lambda\cdot\Eins)$:
\begin{eqnarray}
\label{eq_wfpol_inkl}
  \WFpol{\omega^+}&\supseteq & \bigl\{(x,y;\xi,\eta;a(x,y;\xi,\eta)w)\,|\,
  (x,y;\xi,\eta;w)\in\WFpol{\Lambda\cdot\Eins}\bigr\}  \nonumber\\
  && =\bigl\{(x,y;\xi,\eta;\lambda\cdot\!\xislash)\,|\,
  (x,y;\xi,\eta)\in\WF{\Lambda}, \lambda\in\C\bigr\}. 
\end{eqnarray}
Now $\xislash\neq 0$ if $(x,y;\xi,\eta)\in\WF{\Lambda}$, and since the
projection of the nontrivial part of the polarization set onto the
cotangent bundle gives back the wave front set, we see that by the
action of $A$ the wave front set does not become smaller. Thus
$\omega^+$ has got the same wave front set as the scalar Hadamard
distribution $\Lambda$.

The equality for the polarization set in (\ref{eq_wfpol_inkl}), however,
does not follow from theorem~\ref{thm_wfpol_trans}, since the operator
$A$ is characteristic just in the interesting points: we have $\det
a(x,y;\xi,\eta)=-(\xi^2)^2=0$ for $(x,y;\xi,\eta)\in\WF{\Lambda}$.

Instead, we can give a direct calculation of the polarization vectors
$w$ over a point $(x,y;\xi,\eta)$ in the wave front set.  Because of
(\ref{eq_omegaplus}), such vectors have to be a linear combination of
the unit and gamma matrices, and we can set
$w=\alpha\Eins+\beta_\nu\gamma^\nu$.

By definition, if $a$ is the principal symbol of an operator $A$ with
$A\omega^+=0$, a point $(x,y;\xi,\eta;w)$ can only be contained in the
polarization set of $\omega^+$ if $a(x,y;\xi,\eta)w=0$.  We know such
an operator, since the two point distribution is a solution of the
Dirac equation
\begin{equation}
  (-i\slash{\partial}_x+m)\omega^+(x,y)=0,
\end{equation}
and we obtain
\begin{equation} 
  0=\xislash\cdot w=\xi_\mu\gamma^\mu(\alpha\Eins+\beta_\nu\gamma^\nu)
   =\alpha\xi_\mu\gamma^\mu+\xi_\mu\beta^\mu\Eins
    -i\sum_{\mu>\nu}(\xi_\mu\beta_\nu-\xi_\nu\beta_\mu)\sigma^{\mu\nu},
\end{equation}
where $\sigma^{\mu\nu}=\frac{i}{2}[\gamma^\mu,\gamma^\nu]$.  Hence,
since $\xi\neq 0$, we must have $\alpha=0$ and $\beta=\lambda\cdot\xi$
with $\lambda\in\C$.  This means that the polarization vectors over
points $(x,y;\xi,\eta)\in\WF{\omega^+}$ are indeed only the vectors of
the form $w=\lambda\cdot\!\xislash$.

This completes the proof.
\end{proof}

We are now going to give a generalization to arbitrary globally
hyperbolic spacetimes by using Dencker's theorem on the propagation of
singularities of vector-valued distributions.  First we calculate
Dencker's connection for the Dirac equation:

\begin{thm}
\label{thm_prop_dirac}
Dencker's connection $D_P$ for the Dirac operator $P=(-i\nslash+m)$
acting on the Dirac bundle \DM\/ over an arbitrary spacetime $\Mf$ can
be chosen to coincide with the spin connection lifted to $\pi^*\DM$
and restricted to the null geodesics. The analogous statement holds
for the Dirac operator on the dual bundle. In particular, the
connection is independent of the mass $m$.
\end{thm}

\begin{proof}
  A first (rather involved) proof for the Dirac equation for spinor
  half densities was given by Radzikowski \cite{radzikowski_unp}. Here
  we give a simple proof that holds for the Dirac equation on \DM\@.
  
  Once we haven chosen coordinate frames for the tangent and cotangent
  bundle, the operator
\begin{equation} 
   -i\nslash+m=-i\gamma^\mu(\partial_\mu+\sigma_\mu)+m\Eins
\end{equation}
 has principal and subprincipal symbols
\begin{equation} 
   p(x,\xi)=\xi_\mu\gamma^\mu(x),\quad 
  p^s(x,\xi)=-i\gamma^\mu(x)\sigma_\mu(x)+m\Eins-
       \frac{1}{2i}\partial_\mu\gamma^\mu(x). 
\end{equation}
We choose
\begin{equation} 
   \tilde{p}(x,\xi)=\sqrt{-g(x)}\xi_\mu\gamma^\mu(x),\quad q(x,\xi)
   =\sqrt{-g(x)}g^{\mu\nu}(x)\xi_\mu\xi_\nu, 
\end{equation}
and see that $\tilde{p}p=q\cdot\Eins$, thereby the operator is of real
principal type. The characteristic curves of $q$ are again the null
geodesics.  Dencker's connection acts on spinor fields $w$ along the
bicharacteristics that lie in the kernel of $p(x,\xi)$ for every point
$(x,\xi)$, that is they satisfy $\xislash w(x,\xi)=0$.

Putting together the different terms in Dencker's connection and using
$\xislash w=0$, Hamilton's equations and the properties of the
Christoffel symbols and the Dirac matrices, yields
\begin{eqnarray}
\nonumber
D_Pw&=&H_qw+\frac{1}{2}\{\tilde{p},p\}+i\tilde{p}p^s\\
\nonumber
&=&\frac{d w}{d\tau}+\sqrt{-g}\xi_\nu \left(
   \frac{1}{2}\gamma^\mu(\partial_\mu\gamma^\nu)
   -\frac{1}{2}(\partial_\mu\gamma^\nu)\gamma^\mu
   -\frac{1}{4g}(\partial_\mu g)\gamma^\nu\gamma^\mu\right)w \\
\nonumber 
&&+\sqrt{-g}\xi_\nu\left(\gamma^\nu\gamma^\mu\sigma_\mu
   +im\gamma^\nu-
    \frac{1}{2}\gamma^\nu(\partial_\mu\gamma^\mu)\right)w\\
&=&\left(\frac{d}{d\tau}+\dot{x}^\mu\sigma_\mu\right) w=
(\pi^*\nabla_\tau) w,
\end{eqnarray}
which is just the covariant derivative of $w$ (as a vector field along
the geodesic) along the geodesic line with respect to the spin
connection.

The proof for the adjoint operator is done in the same way.
\end{proof}

The proof of theorem~\ref{thm_dirac} now proceeds like in the scalar
case. The spacetime is deformed such that one part $\hat{U}$ carries
the Minkowski metric.  By the isometry $\rho$, the undeformed part of
the deformed spacetime inherits a spin structure which can be extended
to the whole spacetime, since in the globally hyperbolic case all spin
structures are trivial bundles.  Because the metric stays unchanged in
a neighbourhood of a Cauchy surface, the two point function of the
Hadamard state on the undeformed spacetime induces a Hadamard two
point function on the whole deformed spacetime.

The polarization set in the flat part $\hat{U}$ was calculated above.
On the diagonal we have
\begin{equation} 
   (\hat{x},\hat{x};\hat{\xi},\hat{\eta};w)\in
   \WFpol{\omega^+|_{\hat{U}\times\hat{U}}}\ \Leftrightarrow\ 
   -\hat{\eta}=\hat{\xi}\in\overline{V}^+_{\hat{x}}, \hat{\xi}^2=0;
   w=\lambda\cdot\slash{\hat{\xi}}, \lambda\in\C. 
\end{equation}

In order to calculate the polarization vectors $w$ over a point
$(x,x)\in\tilde{U}\times\tilde{U}$ in the curved part of the
spacetime, we make use of the fact that the two point distribution
solves the equation
\begin{equation} 
  P\omega^+=
  \bigl((-i\nslash+m)\otimes\Eins+\Eins\otimes(i\nslash+m)\bigr)\omega^+=0.
\end{equation}
The curves $(x,x;\xi,-\xi)(\tau)$ such that $x(\tau)$ is a null
geodesic and $\xi(\tau)$ is tangent to the curve $x(\tau)$ for any
$\tau$ are null bicharacteristics of this operator.  As a corollary to
theorem~\ref{thm_prop_dirac}, one can easily see that Dencker's
connection for $P$ is $D_P=\pi^\ast(\nabla_\tau\otimes\nabla_\tau)$
along these curves.

According to theorem~\ref{thm_prop_vect}, the polarization set of
$\omega^+$ consists of Hamilton orbits for the operator $P$. These are
sections $w$ of the bundle $\pi^\ast(\DM\boxtimes\DastM)$ along the
null bicharacteristics that are parallel with respect to the
connection $D_P$.  In our case, for the bicharacteristics
$(x,x,\xi,-\xi)$ this just means that the polarization vectors
$w=\lambda\cdot\xislash$ are parallel transported along the geodesic
curves $(x,x)(\tau)$.  Now we have $\nabla\cdot\xislash=0$ along the
geodesics, since $\xi$ is covariantly constant (the tangent vector of
a geodesic is parallel transported along the curve) as well as the
Dirac matrices.  Thus the polarization vectors retain their form
$w=\lambda\cdot\xislash$, and the polarization set over the diagonal
in the curved part is of the same form as in the flat part of the
spacetime and consists of points $(x,x;\xi,-\xi;\lambda\cdot\xislash)$
such that $\xi$ is lightlike and in the forward light cone, with
arbitrary $\lambda\in\C$.

Because of theorem~\ref{thm_wfpol_trans}, this form of the
polarization set is conserved when the two point distribution is
pulled back to the undeformed spacetime via the isometry $\rho$.

Finally, we obtain the polarization vectors away from the diagonal by
shifting the second spinor to the second point: for the operator
$P=\Eins\otimes(i\nslash+m)$ Dencker's connection is simply
$D_P=\Eins\otimes\nabla$.  Thus a point $(x,y;\xi,-\eta;w)$ is
contained in the polarization set of $\omega^+$ if and only if the
parallel transport along the null geodesic connecting $x$ and $y$,
such that $\xi$ and $\eta$ are tangent to the geodesic, shifts a
polarization vector $\lambda\cdot\xislash$ over the point
$(x,x,\xi,-\xi)$ to $w$.

This is just the statement of theorem~\ref{thm_dirac}.\qed

To summarize, we have proved that, like the wave front set, the
polarization set of the two point function of a Hadamard state is
uniquely fixed by the underlying geometry, and that the fibre over
each point of the wave front set is one-dimensional. Thus the
polarization set takes on the minimal possible form.

It seems now natural to define a Hadamard state by the polarization
set of its two point distribution. A first step in this direction was
undertaken by Hollands \cite{hollands} who takes equation
(\ref{eq_wf_dirac}) for the wave front set, which is a result in our
approach, as the definition of a Hadamard state for the free Dirac
field.  It is then shown that this condition on the wave front set
already fixes the polarization set.


\section{Outlook and conclusion}

In this paper we have demonstrated the usefulness of a microlocal
characterization of the distributions that are of interest in quantum
field theory on curved spacetime. However, this is not confined to the
investigation of quasifree Hadamard states.

In the naive perturbative construction of interacting quantum field
theories one encounters formal products of distributions that are a
priori not well-defined. Over the years, one has learned how to deal
with the divergencies in these formal expressions by different
renormalization techniques, but the understanding of the singularity
structure of the distributions which are involved can lead to some
further insight where these divergencies originate.

For example, in scalar $\phi^4$-theory the perturbative construction
of the $S$-matrix involves the time-ordered two point distribution,
whose formal expansion in terms of Wick products of free fields
contains terms like
\begin{equation}
\label{feyn_prod}
  T_2(x,y)=c_1(i\Delta_F(x,y))^3:\!\phi(x)\phi(y)\!:
    +c_2(i\Delta_F(x,y))^2:\!\phi^2(x)\phi^2(y)\!:+\dots,
\end{equation}
where
\begin{equation}
  \Delta_F=i\Lambda+\Delta^+  
\end{equation}
is the Feynman propagator.  

Because of the singularities of distributions, it is impossible to
define a reasonable product on the whole space of distributions,
especially the products of Feynman propagators in (\ref{feyn_prod}) do
not exist as well-defined distributions. This is the reason for the
well-known ultraviolet divergencies in these expressions.

For certain well-behaved distributions, however, it is possible to
define a product, and it can be shown that a simple condition on the
wave front sets of the factors is sufficient for its existence
\cite{hoermander1}: If the singular directions in the wave front sets
of the two factors over the same point do not add up to zero, the
product can be defined.

The wave front set of the Feynman propagator $\Delta_F$ of a
Hadamard state has also been calculated by Radzikowski
\cite{radzikowski},
\begin{eqnarray}
    \WF{\Delta_F}&=&\bigl\{(x,y;\xi,-\eta)\in 
    \TastMMon\bigr.\,|
      \protect{\begin{array}[t]{l}(x,\xi)\sim (y,\eta),x\not=y,
          \nonumber \\
       \bigl.\xi\in \overline{V}_x^\pm\ \text{if}\
       x\in J^\pm(y)\bigr\} \end{array} } \\
    &&\cup\ \{(x,x;\xi,-\xi)\in\TastMMon\}.
\end{eqnarray} 
The condition $\xi\in \overline{V}_x^\pm$ if $x\in J^\pm(y)$ now ensures
the existence of the products in (\ref{feyn_prod}) at all points away
from the diagonal, while over the points of the diagonal the wave
front sets of the factors are too large. Therefore the only ambiguity
in the definition of the product lies in the extension of the
resulting distribution to the diagonal, and renormalization can be
viewed as the process of extending the time-ordered distributions to
the whole spacetime.

This approach to renormalization theory has not only the advantage of
being mathematically elegant and rigorous, but it also works entirely
in configuration space. Thus it can be extended to a generic
spacetime, where all the powerful renormalization techniques in
momentum space cannot be applied because of the lack of a global
momentum space.  Indeed, in \cite{brun_fred} it could be shown that
scalar field theories on curved spacetimes can be renormalized under
the same conditions as on Minkowski spacetime.

As the main new result of this work, we have calculated the
polarization set for the two point function of a Hadamard state of the
free Dirac field.  The Feynman propagator can be investigated in the
same way, and on Minkowski spacetime it can easily be shown that the
polarization vectors are of the same form as for the two point
function. The generalization to a curved spacetime, however, is
complicated by the fact that the Feynman propagator does not solve the
homogeneous Dirac equation. Therefore Dencker's theorem on the
propagation of singularities can only be used to determine the
polarization vectors away from the diagonal.  Nevertheless, we expect
that the Feynman propagator for physical states does not have a larger
polarization set than on Minkowski spacetime.  The knowledge of the
singularity structure of the Feynman propagator should then contribute
to a better understanding of interacting quantum field theories
containing fermions.


\subsection*{Acknowledgements}

I would like to thank Prof. Klaus Fredenhagen for his guidance in this
work. I am also thankful to Marek Radzikowski who stimulated this
research by helpful discussions and making available his unpublished
results. The opportunity to complete this work under financial support
by DESY is gratefully acknowledged.


\end{document}